\documentclass[
superscriptaddress,
 amsmath,amssymb,
 aps,
 pra,
 floatfix,
twocolumn
]{revtex4-2}

\usepackage{graphicx}
\usepackage{dcolumn}
\usepackage{bm}
\usepackage{bbm}
\usepackage{comment}
\usepackage{physics}
\usepackage[resetlabels,labeled]{multibib}
\usepackage{amsfonts}
\usepackage[dvipsnames]{xcolor}
\usepackage{tikz}
\usetikzlibrary{quantikz}
\usetikzlibrary{external}
\usepackage[ruled,vlined]{algorithm2e}
\usepackage{graphicx}
\usepackage{subfigure}
\usepackage[normalem]{ulem}
\usepackage{amsthm}
\newtheorem{theorem}{Theorem}
\newtheorem{lemma}[theorem]{Lemma}

\newtheorem{definition}[theorem]{Definition}

\newcommand{\aum}[1]{{\color{teal} \textsf{AUM:} #1}} 

\newcommand{\tom}[1]{{\color{magenta} \textsf{TOM:} #1}}
\newcommand{\todo}[1]{{\color{red} \textsf{TODO:} #1}}

\usepackage[breaklinks=true]{hyperref}
\hypersetup{
  colorlinks   = true, 
  urlcolor     = blue, 
  linkcolor    = blue, 
  citecolor   = magenta 
}
\usepackage{soul} 
\usepackage{MnSymbol}
\usepackage{booktabs}

\definecolor{strcyan}{RGB}{33,140,141}

\newcites{sup}{}
\newcommand{\beginsupplement}{%
        \setcounter{equation}{0}
        \setcounter{section}{0}
        \setcounter{figure}{0}
        \setcounter{table}{0}
        \setcounter{page}{1}
        \setcounter{theorem}{0}
        \makeatletter
        \renewcommand{\theequation}{S\arabic{equation}}
        \renewcommand{\thefigure}{S\arabic{figure}}
        \renewcommand{\thetable}{S\arabic{table}}
        \renewcommand{\thesection}{S\arabic{section}}
        \renewcommand{\thetheorem}{S\arabic{theorem}}
        \renewcommand{\thepage}{S\arabic{page}}
        }

\newcommand\av[1]{\left\langle #1 \right\rangle}

\newcommand{\R}{\mathbb R}

\renewcommand{\O}{\mathcal O}
\begin{document}

\title{
   Quantum Next Generation Reservoir Computing: \\ An Efficient Quantum Algorithm for Forecasting Quantum Dynamics
}

\author{Apimuk Sornsaeng}
\affiliation{Chula Intelligent and Complex Systems, Department of Physics, Faculty of Science, Chulalongkorn University, Bangkok, Thailand, 10330}
  
\author{Ninnat Dangniam}
\affiliation{The Institute for Fundamental Study, Naresuan University, Phitsanulok, Thailand, 65000}
\author{Thiparat Chotibut}%
\email[Corresponding author: ]{thiparatc@gmail.com}
\affiliation{Chula Intelligent and Complex Systems, Department of Physics, Faculty of Science, Chulalongkorn University, Bangkok, Thailand, 10330}


\begin{abstract}
Next Generation Reservoir Computing (NG-RC) is a modern class of model-free machine learning that enables an accurate forecasting of time series data generated by dynamical systems. We demonstrate that NG-RC can accurately predict full many-body quantum dynamics in both integrable and chaotic systems. This is in contrast to the conventional application of reservoir computing that concentrates on the prediction of the dynamics of observables. In addition, we apply a technique which we refer to as \emph{skipping ahead} to predict far future states accurately without the need to extract information about the intermediate states.  However, adopting a classical NG-RC for many-body quantum dynamics prediction is computationally prohibitive due to the large Hilbert space of sample input data. In this work, we propose an end-to-end quantum algorithm for many-body quantum dynamics forecasting with a quantum computational speedup via the block-encoding technique. This proposal presents an efficient model-free quantum scheme to forecast quantum dynamics coherently, bypassing inductive biases incurred in a model-based approach.
\end{abstract}



\maketitle


\section{Motivation and Introduction}

 
Learning quantum dynamics presents fundamental challenges in quantum physics, to which numerous machine learning techniques have been applied \cite[and references therein]{huang2023,Caro_natcomm_2023,Siddiqi_PRX2020,troyer_solving_2017, jerbi2023power,Mohseni2022deeplearning,Granade_2012,Rodriguez_2022}.
Significant efforts have been made to compactly represent many-body quantum states and efficiently parameterize their dynamics using classical learning algorithms with classical data \cite{troyer_solving_2017,Mohseni2022deeplearning,Heyl_PRL_2020,Carleo_PRO_2019}
On the other hand, quantum algorithms that work on a large batch of classical data typically requires extreme assumptions about data loading and readout, which remain points of contention, see for example \cite{aaronson2015,tang2019,matteo2020}.

The two challenges spur the proposals of quantum algorithms that can learn directly from quantum data \cite{schtatzki-quantum-data,huang-quantum-data}.  
Our work adds to the literature by providing a novel quantum algorithm for learning from quantum data inspired by next generation reservoir computing (NG-RC) algorithm. Importantly, this model-free paradigm of reservoir computing requires only time-series of quantum data, without assuming a quantum dynamical ansatz {\it a priori}. Such approach circumvents inductive biases that could arise in learning complex quantum dynamics. We commence with a concise overview of the standard RC and the NG-RC, emphasizing the innovative utilization of NG-RC in the accurate prediction of many-body quantum dynamics. Subsequently, we address the limitations inherent to classical NG-RC and introduce a quantum algorithm designed to overcome the challenges presented by its classical counterpart.

{\bf Reservoir computing (RC)}. RC is a computational paradigm in machine learning that harnesses a recurrent neural network (RNN) to learn time-series data, such as the states of a dynamical system. Even when the dynamics is complicated or chaotic, well-tuned RC can accurately forecast the future states of such dynamics up to many Lyapunov times  \cite{pathak2017,pathak2018}.
Importantly, RC bypasses the training of an RNN by utilizing a fixed, randomly initialized RNN, called a \emph{reservoir}, consisting of a large number $L$ of hidden neurons.

Suppose that an input data $\vb*{s}_k\in \mathbb{R}^{D}$ is fed into the reservoir, where $k$ specifies the time step of some dynamical process. The data is represented as a feature vector $\vb*{x}_k = f(\vb*{s}_k,\vb*{x}_{k-1}) \in \R^L$, where $f$ is typically a highly non-linear function that represents the dynamics of the reservoir.
The feature vector is then linearly transformed in the final, trainable output layer into a prediction vector $\vb*{y}'_k = W\vb*{x}_k$. 
In particular, the input $\vb*{s}_k$ at the $k$th time step and the output at the previous time step $\vb*{y}'_{k-1}$ may be fed together as inputs into the reservoir to train the feature vector $\vb*{x}_k$ for the next time step. 


To prevent overfitting in a supervised learning via RC, the weight matrix $W$ is obtained by optimizing $\vb*{y}'_k$ in the least-square sense with respect to the desired target $\vb*{y}_k$, which can be done via the Tikhonov regularization
\begin{equation}\label{eq: W}
    W = YX^T(XX^T+\lambda I)^{-1}.
\end{equation}
Here $Y=(\vb*{y}_0,\ldots,\vb*{y}_{T-1})\in\R^{D\times T}$ is the target matrix, $X = (\vb*{x}_0,\ldots,\vb*{x}_{T-1})\in\R^{L\times T}$ is the feature matrix, and $\lambda\ge0$ is the regularization parameter. Here, $T$ stands for the number of time steps of dynamics used to train the reservoir.


\begin{figure*}[ht]
    \centering
    \resizebox{\textwidth}{!}{
    \includegraphics[scale=1]{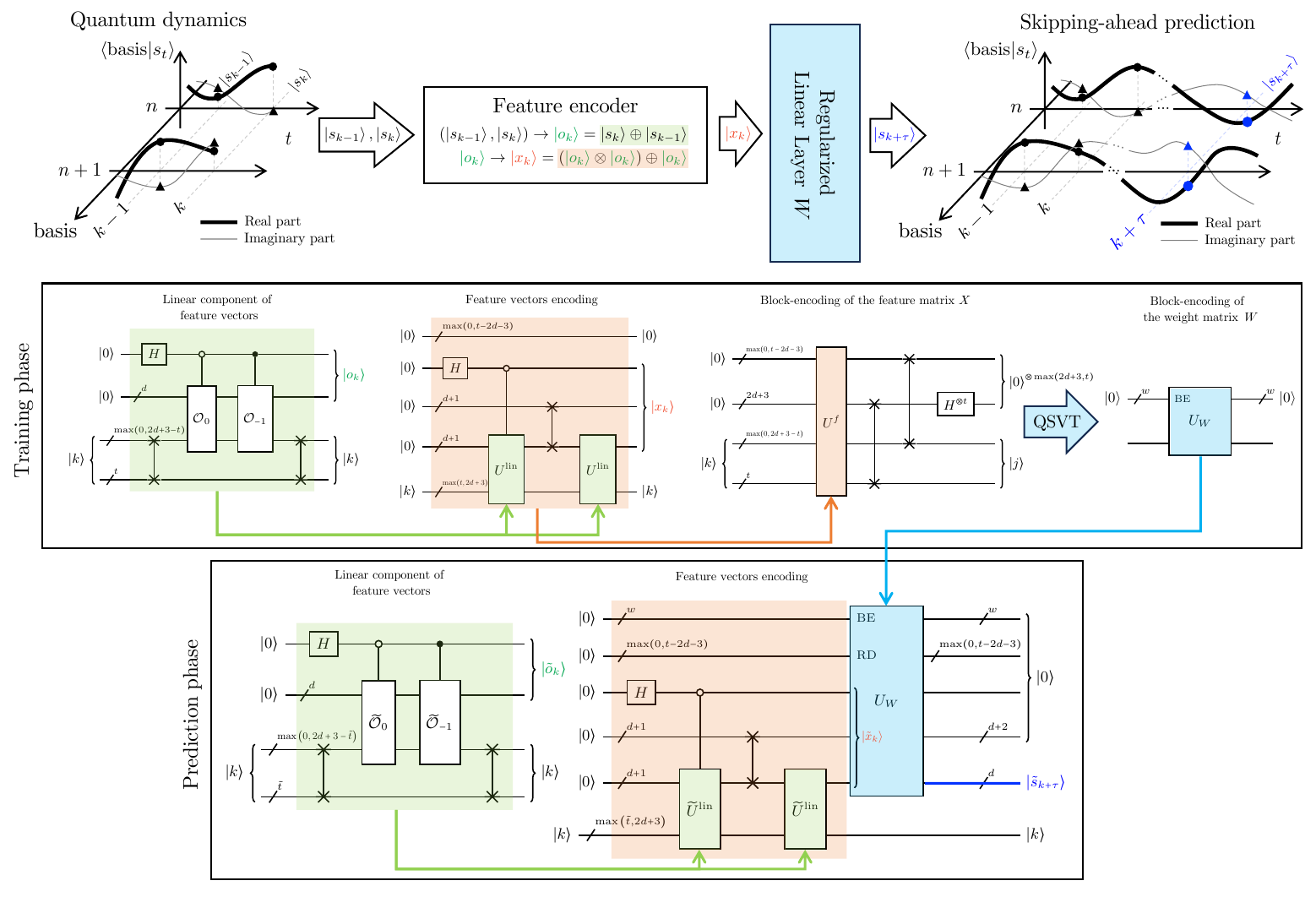}}
    \caption{(Top) The schematic of the QNG-RC algorithm. The initial part of the algorithm encodes the history of quantum dynamics collected from a time series into nonlinear feature vectors $\ket{x_k}$, which is then processed by a regularized linear layer to predict future quantum states. (Middle) In the training phase of QNG-RC, the optimally trained weight matrix in Eq.~\eqref{eq: W} is encoded in a unitary operator and processed by the quantum singular value transform (QSVT), the total complexity of which is summarized in Theorem \ref{thm: W}. (Bottom) In the prediction phase, the feature vector can be constructed via the same quantum circuits as in the training phase. Consequently, the predicted states will be revealed in the blue register in the last quantum circuit after applying the optimal weight matrix from the training phase.}
    \label{fig: roadmap}
\end{figure*}

{\bf NG-RC.} 
Despite the fast training protocol offered by RC, the random initialization of the reservoir presents its own problem: there is an overwhelmingly large number of hyperparameters to be optimized and there is no consensus on how to pick an optimal reservoir. 
NG-RC is an alternative approach that takes advantage of the discovery that RC can be equivalently performed using a \emph{linear} reservoir and a \emph{nonlinear} trainable output layer \cite{gonon2019,hart2021}. The latter is in turn equivalent to a nonlinear vector autoregression (NVAR) machine \cite{bollt2021}.
An NVAR machine predicts future observations of a time series using past observations. In particular, the underlying state space of the dynamics to be learned can be reconstructed
using linear and nonlinear functions of past observations.
Inspired by this correspondence, NG-RC \cite{gauthier2021} proposes 
 taking $m$ time-delay data:
\begin{equation}
\vb*{o}_k=\vb*{s}_{k}\oplus\vb*{s}_{k-\Delta}\oplus\vb*{s}_{k-2\Delta}\oplus\ldots\oplus\vb*{s}_{k-(m-1)\Delta}\label{eq: k-time-delay}
\end{equation}
with step size $\Delta$ as an input, and, forgoing the need for a reservoir, directly constructing a feature vector, 
\begin{align}
    \vb*{x}_k=\vb*{o}_k\oplus(\vb*{o}_k)^{\otimes p},\label{eq: degree-p}
\end{align}
whose nonlinearity arises from a degree-$p$ monomial of the $m$-delay data, $\Delta$, $m$, and $p$ being the NG-RC hyperparameters. 
The feature vector is then optimized via a linear least-square regularization to predict the target dynamics.

NG-RC with $\Delta=1$, $m=2$, and $p=2$ has been used to efficiently predict the dynamics of the Lorenz attractor using only small data sets \cite{gauthier2021}.
The feature vector in this case is a $(4D^2+2D)$-dimensional vector of the form
\begin{align}\label{eq: feature_vec}
    \vb*{x}_k = \vb*{s}_{k}\oplus\vb*{s}_{k-1} \oplus [(\vb*{s}_{k}\oplus\vb*{s}_{k-1}) \otimes (\vb*{s}_{k}\oplus\vb*{s}_{k-1})].
\end{align}

While NG-RC has been demonstrated to effectively forecast complex classical dynamics such as spatiotemporal chaos \cite{ng-rc_spatiotemporalchaos_gauthier} and exhibits robust predictive capabilities despite noisy input dynamics \cite{Liu_noisy_ng-rc_2023}, its potential application to forecasting the dynamics of quantum observables and more generally of full many-body quantum states, which encompass both real and imaginary components, remains unexplored. 

\begin{figure*}[ht]
    \resizebox{\linewidth}{!}{
    \centering
    \includegraphics[scale=1]{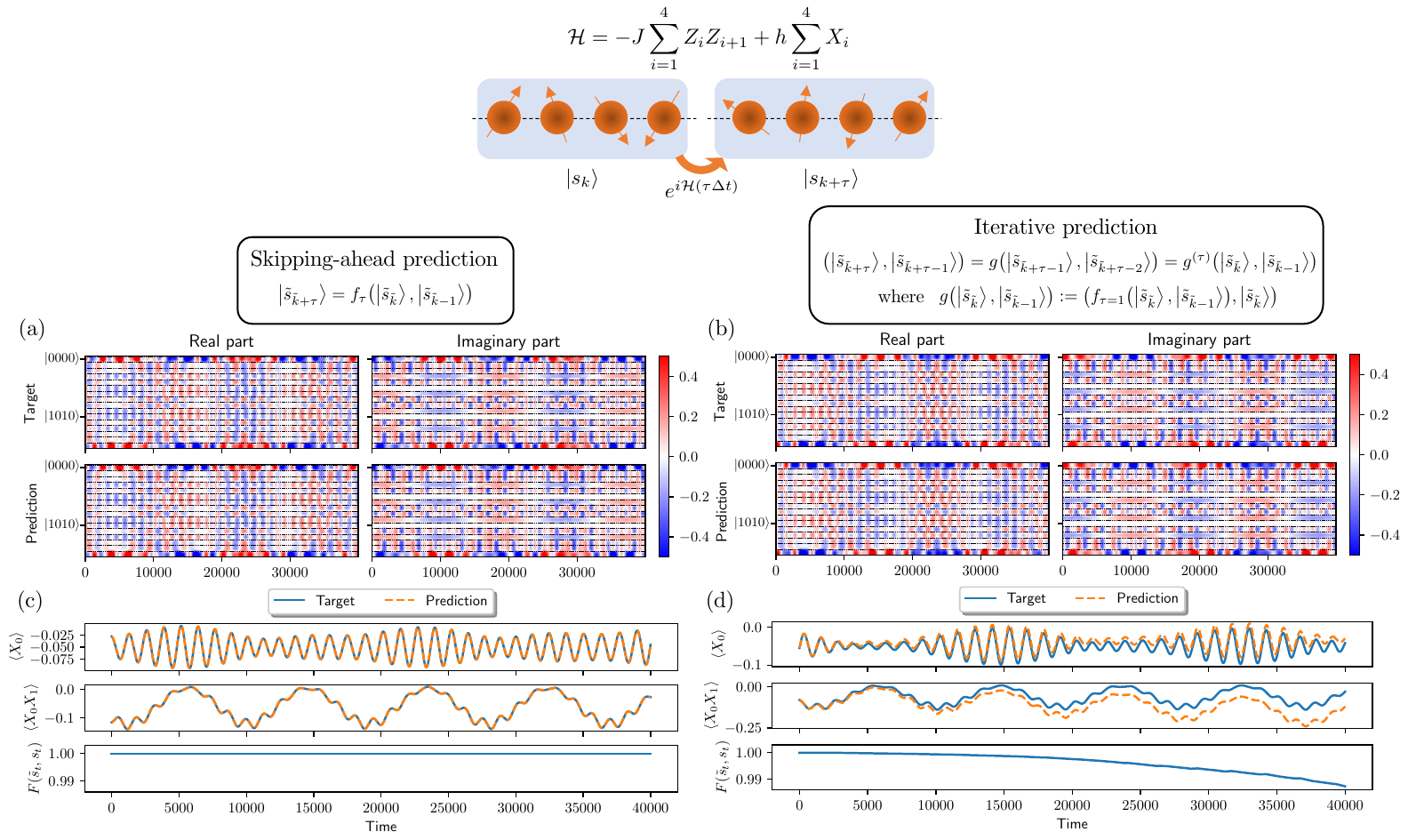}}
    \caption{
    The performances of the NG-RC in predicting {\it unseen} future states of a four-qubit transverse-field Ising model in the disordered phase with $J=0.5$ and $h=5$ across $\widetilde T = 4 \times 10^4$ future time steps employing two different approaches: the skipping-ahead method with a time skip of $\tau=10^6$ steps (the map $f_\tau$ represents the trained NG-RC for predicting the next $\tau$ step from the current step), and the conventional approach of iteratively predicting each successive time step ($g^{(\tau)}$ is the composition of $\tau$ successive $g$ obtained from the trained NG-RC $f_{\tau=1}$.
    $T=2\times 10^4$ steps of the time evolution with the same step size of $\Delta t=1/(200E_{\textrm{max}})$, where $E_{\max}$ is the highest eigen-energy of the system, are used to train the NG-RC in both cases. The training is optimized with the optimal regularization parameter $\lambda = 1.0\times 10^{-3}$ in the iterative prediction and with $\lambda = 0$ in the skipping-ahead prediction. Our benchmark targets are the real and the imaginary parts of all amplitudes in the computational basis ((a) and (b)). The comparisons of the expectation values $\expval{X_0}$, $\expval{X_0X_1}$, and the fidelity $F(\tilde{s}_t,s_t):=|\braket{\tilde{s}_t}{s_t}|$ between the target and the predicted states are shown in (c) and (d).}
    \label{fig: QNGRC_sim}
\end{figure*}



{\bf Learning quantum dynamics with NG-RC.}
Here we demonstrate the capability of NG-RC to predict a unitary quantum dynamics using the strategy we refer to as \emph{skipping ahead}.
Instead of training the output of the feature vector \eqref{eq: feature_vec} to match the target state at the next time step, $\vb*{y}_k = \vb*{s}_{k+1}$, we could train it to match the target state at the next $\tau$ steps, skipping into the future. That is, given $\{\vb*x_k\}_{k=0}^{T-1}$ according to Eq.~\eqref{eq: feature_vec}, the next generation reservoir is trained using $\{\vb*{y}_k = \vb*s_{k+\tau}\}_{k=0}^{T-1}$.
We found that 
our approach of skipping ahead yields a remarkably accurate prediction of the full quantum state into the far future, retaining the fidelity measure close to 1 in the standard example of \emph{integrable} one-dimensional many-body dynamics that we now discuss.    

The one-dimensional transverse-field Ising model is governed by the Hamiltonian
\begin{equation*}
    \mathcal{H}=-J\sum_{i=1}^d Z_{i}Z_{i+1} + h\sum_{i=1}^d X_{i},
\end{equation*}
with a periodic boundary condition, where $X_i$ (resp. $Z_i$) is the Pauli-X (resp. Z) operator acting on the $i$th qubit. The system can be exactly solved by mapping to the free-fermionic model via the Jordan-Wigner transformation. However, the computation of certain expectation values such as the two-point correlation function $\av{X_iX_j}$, let alone the computation of the time-evolved state, on a classical computer becomes infeasible as the number of qubits becomes large \cite{bravyi2002fermionic,beenakker2004charge}. (Note that the model can be simulated efficiently on a quantum computer by exact diagonalization \cite{cervera2018exact}.) 

We benchmark the prediction for the time evolution of a four-qubit transverse-field Ising model in the disordered phase with $J=0.5$ and $h=5$.  The initial quantum state $\ket{0000}$ followed by a long burn-in period initializes the inputs for the predictions of unseen $\widetilde{T}=4\times 10^4$ time steps in the future. (After the burn-in period, the input states are far into the future of those from the training set.)  Fig. \ref{fig: QNGRC_sim} (left column) shows the results employing the skipping-ahead strategy with the time skip of $\tau=10^6$ steps into the future, whereas Fig. \ref{fig: QNGRC_sim} (right column) shows the results from the conventional iterative NG-RC approach to predict successive states in the future. Notably, the skipping-ahead strategy significantly outperforms the conventional approach as reflected by the almost perfect fidelity between the predicted states and the target states, as well as the almost perfect prediction of the observables. The iterative approach's performances drop as the number of iteration increases, as expected.

In addition, we also benchmark our skipping-ahead approach with a \emph{non-integrable}, \emph{chaotic} many-body dynamics in a tilted-field Ising model with five qubits. The fidelity of the predicted states remain above 0.99, showing an excellent agreement with the full many-body states into the future. The details and results are provided in Sec. \ref{sec: Sup-TiltedField}.

{\bf Challenges in using NG-RC for quantum dynamics predictions.} The preceding results provide a piece of evidence that NG-RC may offer an alternative route to a high-precision prediction of quantum dynamics. However, the primary computational bottleneck arises from matrix inversion in Eq.~\eqref{eq: W} (or its generalization for complex-valued data with $X^T$ replaced by $X^{\dagger}$). The Tikhonov-regularized least squares requires $O(MN^2)$ matrix operations, where $N$ and $M$ are,  respectively, the larger and the smaller dimension of the matrix. This complexity is computationally prohibitive  when the data matrix is collected from many-body quantum states, since the dimension of each column vector scales exponentially with the system size. In the following section, we present an end-to-end, \textit{quantum next generation reservoir computing} (QNG-RC) algorithm that does not suffer from the exponential complexity of the classical counterpart. Our algorithm takes as inputs quantum data and systematically produce coherent predictions of future quantum states as outputs by utilizing block-encoding techniques and the quantum singular value transform (QSVT) \cite{gilyen2019,Martyn2021}, which enables efficient implementations of linear-algebraic operations at the root of NG-RC.

{\bf Related works.} 
A significant portion of work on quantum variants of classical reservoir computing, referred to as  \textit{quantum reservoir computing} (QRC), utilizes quantum systems as   reservoirs (substrates) \cite{fujii2017RC,zambrini2020RC,zambrini2021gaussian,Mujal2021,Fujii2021, Kutvonen_Fujii_2020, zambrini_PRL_2021,Mujal_Zambrini_NPJ_QI_2023,Ortega_PRE_2023, Nakajima_PRR_2023,fry_chen_2023}. However, the outputs of these QRCs remain classical in nature, namely the expectation values of some observables, which are subsequently used to predict classical dynamical systems, not the full quantum state dynamics as in this work.
Other QRC proposals focus on the task to extract partial information about quantum states such as their entanglement or purity \cite{ghosh2019QRC}, or perform a complete tomography \cite{ghosh2020reconstruct}. 
Ref. \cite{kawai2020excited_states} proposes learning excited states of a Hamiltonian by measuring expectation values in the ground state, which are then subsequently optimized using classical machine learning techniques. In all three examples, the main goal is to use QRC to effectively transform final single-qubit measurements into entangled ones that contains the desired information about the quantum state of interest, thus reducing the number or the complexity of measurement settings.

Nevertheless, there exist notable works that share several characteristics with ours. Carleo and Troyer \cite{troyer_solving_2017} introduces a method for learning quantum states and dynamics using neural network states, circumventing inductive biases and the exponential size of classical memory needed to process quantum data. Since reservoirs can be considered as special cases of recurrent neural networks, conventional QRC cannot surpass the predictions of the best quantum neural networks. However, the training of neural networks itself is a complex task in its own right, and both RC and NG-RC offers a means to navigate this challenge. 
Moreover, Ref. \cite{troyer_solving_2017} learns a quantum dynamics by classically processing the data obtained by sampling from the trained network, a point that diverges from the approach presented in our work. 
Another related contribution, highlighted in Ref. \cite{ghosh2021circuit} by Ghosh \emph{et al.}, employs QRC to coherently simulate the target state of a quantum dynamics, whereas our approach shifts the methodology to that of a next generation RC. 

While our skipping-ahead method offers a fast route to generate quantum states that would otherwise emerge in the far future through the natural progression of quantum dynamics, the technique should not be confused with the fast-forwarding of quantum evolution, since our method requires training samples from the equally distant past. Moreover, our prediction operates in a model-free framework, devoid of Trotterization techniques. This distinction ensures that our proposed skipping-ahead approach is fundamentally different from fast forwarding.  Consequently, the constraints outlined in the no-fast-forwarding literature \cite{Cirstoiu_Holmes_2020, Berry_no-ff_2007, Atia_no-ff_2017, no-ff_IEEE,Gu2021fastforwarding} should not apply to our prediction method. 

\section{Method and Results}

Here we propose an efficient QNG-RC algorithm, utilizing the block-encoding technology for matrix manipulations with the QSVT.

{\bf Block encoding.} 
We employ the block-encoding technique to construct the non-linear feature matrix and perform the Tikhonov regularization. Specifically, a relevant matrix $A$ is embedded  as a submatrix of a unitary gate $U$ such that
\begin{align}
     \left(\bra{0}^{\otimes a}\otimes I^{\otimes s}\right) U \left(\ket{0}^{\otimes a}\otimes I^{\otimes s}\right) = A/\alpha \text{ for } U = \mqty(A&*\\ *&*),
\end{align}
where $\ket{0}^{\otimes a}$ is the fiducial state of an $a$-qubit ancillary system, and $\alpha \ge \norm{A}$ due to the unitary constraint. Def. \ref{def: block_main} is a concrete definition of the block-encoding.
{\definition \label{def: block_main}
Suppose that $A$ is an $s$-qubit operator, the parameters $\alpha,\epsilon\in\mathbb{R}$, and $a\in\mathbb{N}$. The $(\alpha,a,\epsilon)$-block-encoding of a $(s+a)$-qubit unitary operator $U$ can be defined if
\begin{equation*}
    \norm{A-\alpha\qty(\bra{0}^{\otimes a}\otimes I)U\qty(\ket{0}^{\otimes a}\otimes I)}\leq\epsilon.
\end{equation*}}

Once the block-encoded matrix is in place,
QSVT allows us to construct a degree-$q$ polynomial approximation of essentially any well-behaved function of the singular values of $A$, using the number of gates $U$ and controlled operations that are efficient in $q$ \cite{gilyen2019,Martyn2021}.
This approach enables straightforward creation of matrix polynomials, and in particular, creation of the Moore-Penrose pseudo-inverse by inverting the singular values of $A$. 
Note that the Tikhonov regularization in the block-encoded form is also presented as a subroutine in Ref. \cite{Chakraborty2023}. Our approach here, outlined in Lemma \ref{lemma: pinv_lambda} in the supplementary material, is slightly distinct from theirs.

{\bf Input and target assumptions.} The forecasting of quantum dynamics via QNG-RC is divided into two phases: the training phase and the prediction phase.  
The data are assumed to be given by the 
oracles
\begin{subequations}
\begin{align}
    \O_0: \ket{0}^{\otimes d}\ket{k} &\mapsto \ket{s_k}\ket{k},\\
    \O_{-1}: \ket{0}^{\otimes d}\ket{k} &\mapsto \ket{s_{k-1}}\ket{k}, \\
    \O_{\tau}: \ket{0}^{\otimes d}\ket{k} &\mapsto \ket{s_{k+\tau}}\ket{k}, \\
    \widetilde{\mathcal{O}}_0: \ket{0}^{\otimes d}\ket{\tilde{k}} &\mapsto \ket{\tilde{s}_{\tilde{k}}}\ket{\tilde{k}}, \label{eq:oracle-tilde0}\\
    \widetilde{\mathcal{O}}_{-1}: \ket{0}^{\otimes d}\ket{\tilde{k}} &\mapsto \ket{\tilde{s}_{\tilde{k}-1}}\ket{\tilde{k}}, \label{eq:oracle-tilde1}
\end{align}
\end{subequations}
where $\ket{s_k}$ and $\ket{\tilde{s}_{\tilde{k}}}$ are $d$-qubit input data in the training phase and in the prediction phase respectively, and $k=0,\ldots,T-1$ and $\tilde{k}=0,\ldots,\widetilde{T}-1$.
This assumption is equivalent to an access to the controlled version of a unitary that generates each data point, a common assumption in the block-encoding literature. 
In particular, we must be able to create coherent superpositions of the form $\ket{o_k}=(\ket{0}\ket{s_{k}}+\ket{1}\ket{s_{k-1}})/\sqrt{2}$. 

To express the total query complexity of the algorithm, we denote the number of calls to the oracle $\O_i$ (resp.  $\widetilde{\O}_i$) by $T_{\O}$ (resp. $T_{\widetilde \O}$). For convenience, the states created by the oracles are assumed to be error-free, but the total query complexity will take into account errors that occur in the block encoding processes of various matrices.

{\bf Training phase}
According to the NG-RC procedure, by utilizing regularized linear optimization, we obtain an optimal weight matrix through the Tikhonov-regularized pseudoinverse of the feature matrix $X$ (cf. Eq.~\eqref{eq: W} with complex elements). Therefore, we begin by efficiently constructing the feature matrix $X$. Assuming the existence of oracles $\O_0$ and $\O_{-1}$, we can generate the linear component of the feature vector, denoted as $\ket{o_k}$, by introducing an additional qubit to entangle with these oracles. This process is depicted by the operator $U^{\text{lin}}$, shown as a green box in Fig. \ref{fig: roadmap}, which maps the state $\ket{0}^{\otimes d+1}\ket{k}$ to $\ket{o_k}\ket{k}$. To incorporate the nonlinear component $\ket{o_k}\otimes\ket{o_k}$ into the feature vector, we must apply the operator $U^{\text{lin}}$ twice due to the constraints imposed by the no-cloning theorem. The resulting feature vector $\ket{x_k}$ \eqref{eq: feature_vec} is represented by the quantum circuit $U^f$, which maps the state $\ket{0}^{\otimes 2d+3}\ket{k}$ to $\ket{x_k}\ket{k}$, as also illustrated in Fig. \ref{fig: roadmap} (middle, orange box).
Note that, with this encoding method, the dimension of $\ket{x_k}$ is $8D^2$, wherein $4D^2+2D$ dimensions constitute the feature vector $\vb*{x}_k$ analogous to the case of classical NG-RC \eqref{eq: feature_vec}, and the remaining dimensions are for zero padding. We can coherently construct the feature matrix $X=\mqty(\ket{x_0},\cdots,\ket{x_{T-1}})\in\mathbb{C}^{8D^2\times T}$ with the block-encoding technology using quantum gates 
\begin{equation*}
    \qty(I^{\otimes \max(0,2d+3-t)}\otimes H^{\otimes t}\otimes I^{\otimes \max(2d+3,t)})\cdot \text{\sc SWAP}\cdot U^f, 
\end{equation*}
which is the $(\sqrt{T},\max(2d+3,t),0)$-block-encoding of the feature matrix $X$, illustrated in Fig. \ref{fig: roadmap} (middle). Note that the dimension of $X$ here is $\max(T,8D^2)\times \max(T,8D^2)$, which includes another zero padding to convert the submatrix $X$ in the block-encoding unitary into a square submatrix \cite{gilyen2019}.

Due to the requirement for matrix multiplication involving $Y$ and the pseudoinverse of $X$ in Eq.~\eqref{eq: W}, the size of submatrix $Y$ in the block-encoding unitary must align with the dimension of the submatrix $X$. Utilizing block encoding techniques, the quantum circuit 
\begin{equation*}
\qty(I^{\otimes \max(0,2d+3-t)}\otimes H^{\otimes t}\otimes I^{\otimes \max(2d+3,t)})\cdot \textsc{SWAP}\cdot \mathcal{O}_\tau,
\end{equation*}
along with pre-amplification proposed in Ref.~\cite{power_block}, can effectively create the $(\sqrt{2}\|Y\|,\max(2d+3,t)+1,\delta_Y)$-block encoding of $Y$, where $\delta_Y$ is the error of the pre-amplification.


By virtue of Lemma \ref{lemma: pinv_lambda} in the supplementary material, the optimal regularized weight matrix~\eqref{eq: W} in the block-encoded form can be constructed with the resources stated below in Theorem \ref{thm: W}.


{\theorem{\label{thm: W}
Let $\delta_W\in(0,1]$. Suppose that we have the $(\sqrt{T},\max(2d+3,t),0)$-block-encoding of the feature matrix $X$ and the $(\sqrt{2}\|Y\|,\max(2d+3,t)+1,\delta_Y)$-block encoding of training target matrix $Y$. Let $\kappa_X$ be the condition number of $X$, and
\begin{equation}\label{eq: kappa}
        \kappa = \kappa_X\sqrt{\frac{\norm{X}^2+\lambda}{\norm{X}^2+\lambda\kappa_X^2}}.
\end{equation}
where $\lambda \geq 0$ is the regularization parameter \cite{Chakraborty2023}. 
Also define
\begin{equation}
    w=2\max(2d+3,t)+2
\end{equation}
to be the number of ancilla qubits used in the block encoding of $W$. In a condition such that $\delta_Y\leq \delta_W/(4\kappa)$, then we can construct the unitary operator $U_W$ as a $\qty(\frac{2\sqrt{2}\kappa\|Y\|}{\norm{X}+\sqrt{\lambda}},w,\delta_W)$-block-encoding of the weight matrix $W$ in 
\begin{equation}\label{eq: query-complexity-W-norm}
T_W=O\qty(\qty(\frac{\kappa}{\norm{X}+\sqrt{\lambda}}+\frac{1}{\norm{Y}})\log\qty(\frac{\kappa\norm{Y}}{\delta_W})T_\O\sqrt{T})
\end{equation}
queries, where $T_\O$ is the number of calls to the oracles.}}

{In the case $t>2d+3$, $U_W$ would be a $\qty(\frac{2\sqrt{2}\kappa\|Y\|}{\norm{X}+\sqrt{\lambda}},2t+2,\delta_W)$-block encoding of the weight matrix $W$.}
The query complexity~\eqref{eq: query-complexity-W-norm} can be expressed more directly in terms of $T$ and $D$ using inequality of operator norm. Since each column in both $X$ and $Y$ are unit vector, implying that we have an inequality $\frac{1}{\sqrt{T}}\leq\norm{X}\leq D\sqrt{T}$ and $\frac{1}{\sqrt{T}}\leq\norm{Y}\leq \sqrt{DT}$, respectively. Matrix norm inequalities are detailed in Sec. \ref{subsec: matrix_norm} of the supplementary material. By assuming $\lambda=O(1)$, the time complexity is expressed to
\begin{equation}
    T_W = O\qty(\kappa T\log\qty(\frac{\kappa DT}{\delta_W})T_\O),
\end{equation}
Note that the time complexity here is difficult to compare to that of the classical NG-RC because of the unknown gate complexity of implementing $\O_i$. We can only say that the number of calls to $\O_i$ is of the order of $O\qty(\kappa T\log\qty(\frac{\kappa DT}{\delta_W}))$.


{\bf Prediction phase.} Now that we have the trained weight matrix in the block-encoded form, $U_W$, we can build a circuit to predict future quantum states by the skipping-ahead method. 
As shown in Fig. \ref{fig: roadmap},
the quantum circuit for the prediction phase of the QNG-RC has the same structure as that for the training phase, but with input past quantum states incorporated via the oracles $\widetilde{\mathcal{O}}_0$ and $\widetilde{\mathcal{O}}_{-1}$ defined in Eqs.~\eqref{eq:oracle-tilde0} and~\eqref{eq:oracle-tilde1}, respectively.
The output of the circuit is the desired prediction vector $\ket{\tilde{s}_{\tilde{k}+\tau}}=W\ket{\tilde{x}_{\tilde{k}}}/\norm{W\ket{\tilde{x}_{\tilde{k}}}}_2$ up to an error $\delta\in (0,1)$ due to an inaccuracy in the application of the weight matrix $W$. More specifically, we have an inequality
\begin{equation}
    \delta_W\leq \frac{\delta\norm{W}}{4\kappa_W},
\end{equation}
where $\delta_W$ is the error in Theorem \ref{thm: W}, and $\kappa_W$ is the condition number of $W$. 
Employing the pre-amplification technique for block encoding applied on a quantum state in Ref. \cite{Chakraborty2023}, the number of queries to produce such a $\delta$-close prediction vector is
\begin{equation*}
O\qty(\frac{\kappa\kappa_W}{\norm{X}+\sqrt{\lambda}}\frac{\norm{Y}}{\norm{W}}\log\qty(\frac{\kappa_W}{\delta})T_W+\kappa_WT_{\widetilde \O}), 
\end{equation*}
where $\kappa$ is defined in Eq.~\eqref{eq: kappa}. What is more, the operator norm inequality allows us to rewrite the query complexity as $O\qty({\kappa\kappa_W}\log\qty(\kappa_W/\delta)T_W+\kappa_WT_{\widetilde \O})$. The technical details regarding the pre-amplification and the operator norm inequality are given in Lemma \ref{lemma: operation_state} and Sec. \ref{subsec: matrix_norm} in the supplementary material.

\emph{Number of ancilla qubits.}--The overall circuit for the prediction represents the unitary operator
\begin{equation}
    \tilde{U}: \ket{0}^{\otimes w'}\ket{0}^{\otimes 2d+3}\ket{\tilde{k}}\mapsto\ket{0}^{\otimes (w'+d+3)}\ket{\tilde{s}_{\tilde{k}+\tau}}\ket{\tilde{k}},
\end{equation}
where the register $\ket{\tilde{k}}$ contains $\max\qty(\tilde{t},2d+3)$ qubits. The block encoding requires in total $w'=w+\max(0,t-2d-3)$ qubits. The $w$ given in  Theorem \ref{thm: W}, is the number of index qubits in the block-encoding (BE) register, and $\max(0,t-2d-3)$ is the number of qubits used in the zero padding in the register RD, see Fig. \ref{fig: roadmap} and the corresponding caption. 
In the case $t>2d+3$, the number of ancilla qubits is $w'=3t-2d-1$ and $w'=4d+8$ for $t<2d+3$.

\section{Conclusion and Outlooks}

Drawing inspiration from the paradigm of NG-RC, we develop a novel, end-to-end quantum algorithm for predicting \textit{skipped-ahead} many-body quantum dynamics. The algorithm is purely data-driven, only requiring a time-series of quantum data and no assumption is made on the nature of the dynamics i.e. the forecasting is \emph{model-free}.
The algorithm employs block encoding to efficiently handle matrix algebra subroutines, including matrix multiplication, inversion, and regularized linear optimization, which are manipulated by the QSVT. In addition to providing a quantum computational speedup and avoiding exponential resource consumption in storing many-body quantum states classically, our algorithm coherently processes and generates quantum data, thereby circumventing classical-quantum data conversion problems during encoding and readout procedures.

In the original NG-RC framework, a feature vector $\vb*{x}_i$ can be constructed with a degree-$p$ monomial of an $m$-time-delay data, see Eqs.~\eqref{eq: k-time-delay} - \eqref{eq: degree-p}. In our quantum circuit design, though we assumed $m=2$ for simplicity, the quantum operator $U^\text{lin}$ can be slightly modified to incorporate the oracles $\qty{\mathcal{O}_{-j\Delta}}_{j=0}^{m-1}$ for an $m$-time-delayed $\ket{o_k}$. A quantum circuit that encodes the nonlinear map of Fig. \ref{fig: roadmap} would then involve the addition of multiple $U^\text{lin}$ and more qubits to accommodate the tensor product $(\vb*{o}_i)^{\otimes p}$. The construction of the corresponding quantum circuits of this more general form of non-linearity is elaborated in Sec. \ref{sec: time_delay} of the supplementary material.


Lastly, the conventional application of NG-RC to make iterative predictions, whose results are shown in Fig. \ref{fig: QNGRC_sim}(b), can also be implemented on a quantum circuit. However, the number of gates and the error scales exponentially with the number of iterations. 
Nevertheless, we discuss a quantum circuit for this approach in  Sec. \ref{sec: iterative} of the supplementary material, along with the analysis of the error propagation.



\section*{Acknowledgement}
We thank Supanut Thanasilp, Zo\"{e} Holmes, and Dario Poletti for a helpful discussion. This research is supported by the Program Management Unit for Human Resources and Institutional Development, Research and Innovation (Grant No. B05F650024) and by Thailand Science research and Innovation Fund Chulalongkorn University (IND66230005). The authors acknowledge the National Science and Technology Development Agency, National e-Science Infrastructure Consortium, Chulalongkorn University and the Chulalongkorn Academic Advancement into Its 2nd Century Project, NSRF via the Program Management Unit for Human Resources \& Institutional Development, Research and Innovation [grant numbers B05F650021, B37G660013] (Thailand) for providing computing infrastructure that has contributed to the research results within this paper.

\bibliography{ref}

\beginsupplement

\clearpage
\onecolumngrid
\begin{center}
    \large\textbf{Supplementary Material\\
    Quantum Next Generation Reservoir Computing: \\ An Efficient Quantum Algorithm for Forecasting Quantum Dynamics}
\end{center}
\maketitle
\onecolumngrid

\section{Matrix manipulations with block encoding}
In our proposed algorithm, we construct the quantum counterpart of the NG-RC via the block-encoding formalism. A unitary matrix $U$ is said to be a block encoding of the matrix $A$ if it is in the form
\begin{equation}
    U=\mqty(A/\alpha&\cdot\\\cdot&\cdot)\Longleftrightarrow A=\alpha\qty(\bra{0}^{\otimes a}\otimes I)U\qty(\ket{0}^{\otimes a}\otimes I),
\end{equation}
where $\alpha > \norm{A}$ due to the unitary constraint. The upper-left block of $U$ is labeled by index qubits being in the state $\ket{0}^{\otimes n}$, a projective measurement onto which is required to successfully apply the correct block to the quantum state. 
Formally, we can define a block encoding of the matrix as follows.
\begin{definition}[Block-encoding]\label{def: block}
Suppose that $A$ is an $s$-qubit operator, the parameters $\alpha,\epsilon\in\mathbb{R}$, and $a\in\mathbb{N}$. An $(s+a)$-qubit unitary operator $U$ is an $(\alpha,a,\epsilon)$-block encoding of $A$ if
\begin{equation*}
    \norm{A-\alpha\qty(\bra{0}^{\otimes a}\otimes I)U\qty(\ket{0}^{\otimes a}\otimes I)}\leq\epsilon.
\end{equation*}
\end{definition}
\noindent Note that $\|\cdot\|$ is the spectral norm, which is the largest singular value of a matrix. Techniques are readily available to perform algebraic operations such as matrix addition, matrix multiplication, and tensor product of two block-encoded matrices. 
Matrix multiplication subroutines, for example, which we need to perform the Tikhonov regularization, can be implemented via the following lemma.

\begin{lemma}[Multiplication of two block-encoded matrices {{\cite[Lemma 53]{gilyen2019}}}]\label{lemma: mult}
If $U_A$ and $U_B$ are a $(\alpha,a,\epsilon_A)$-block-encoding of an $s$-qubit operator $A$ and a $(\beta,b,\epsilon_B)$-block-encoding of an $s$-qubit operator $B$, prepared in time $O(T_A)$ and $O(T_B)$, respectively, then $(I_b\otimes U_A)(I_a\otimes U_B)$ is an $(\alpha\beta,a+b,\alpha\epsilon_B+\beta\epsilon_A)$-block encoding of $AB$, which can be prepared in time $O(T_A+T_B)$.
\end{lemma}
\textit{Proof.} See Lemma 53 in Ref. \cite{gilyen2019}.



The probability of projecting the index qubits to the right state depends onto the subnormalization constant $\alpha$. The following lemma, proven in \cite{power_block}, amplifies the success probability by essentially reducing $\alpha$ to the spectral norm of the encoded matrix, thus improving the computational complexity of all block-encoding subroutines.

\begin{lemma}[Pre-amplification of block encoding]\label{lemma: preamplify}
Let $A\in C^{M\times N}$ and $\delta\in(0,1]$. Suppose $U$ is a $(\alpha,a,\epsilon)$-block-encoding of $A$ such that $\epsilon\leq\delta/2$, that can be implemented at a cost $T_U$. Then a $(\sqrt{2}\|A\|,a+1,\delta)$-block-encoding of $A$ can be implemented at a cost of $O\qty(\frac{\alpha}{\|A\|}\log\qty(\frac{\|A\|}{\delta})T_U)$.
\end{lemma}
\textit{Proof.} See Theorem 30 in Ref. \cite{power_block}.

Such pre-amplified block-encoded matrix can be applied to a quantum state efficiently, where the time complexity depends on the error $\epsilon$ of the block encoding as stated in the following lemma.

\begin{lemma}[Applying a pre-amplified block-encoded matrix on a quantum state {{\cite[Corollary 14]{Chakraborty2023}}}]\label{lemma: operation_state} Let $A$ be a quantum operator acting on $s$ qubits, with its singular values confined to the interval $[\norm{A}/\kappa,\norm{A}]$. Consider $\delta\in (0,1)$, and let $U_A$ constitute an $(\alpha,a,\epsilon)$-block-encoding of $A$ such that $\epsilon\leq\frac{\delta\norm{A}}{4\kappa}$; this encoding can be accomplished within a time span of $T_A$. Additionally, assume that $\ket{b}$ represents a quantum state of $s$ qubits that can be prepared in $T_b$ time.

Under these conditions, it is feasible to create a quantum state that approximates $A\ket{b}/\norm{A\ket{b}}$ with a precision of $\delta$, achieving a success probability of at least $\Omega(1)$. This preparation can be executed at a cost of: 
\begin{equation*}
    O\qty(\frac{\alpha\kappa}{\norm{A}}\log\qty(\frac{\kappa}{\delta})T_A+\kappa T_b).
\end{equation*}
\end{lemma}

Matrix (pseudo)inversion serves as a key subroutine of numerous quantum algorithms, including ours. in the QNG-RC algorithm, the Tikhonov regularization centers around the computation of the Moore-Penrose pseudoinverse. The pseudoinverse $A^+$ of $A$ can be constructed in the block-encoded form utilizing the quantum singular-value transform (QSVT). The QSVT framework allows one to transform a (possibly non-square) block-encoded matrix $A$ with singular values $\{\sigma_j\}_j$ to a matrix with the same left- and right-singular vectors, but with the new singular values $\{f(\sigma_j)\}_j$, where $f$ is essentially any polynomial modulo certain technical restrictions. The complexity of the transformation is linear in the degree of the polynomial.
Ref. \cite{Martyn2021} proposed a low-degree polynomial for approximately inverting a matrix in the QSVT procedure, summarized here in Lemma \ref{lemma: inv_poly}, and its QSVT procedure stated in Theorem \ref{thm: inv}.
\begin{lemma}[QSVT polynomial for matrix inversion]\label{lemma: inv_poly}
Given $\kappa \geq 1$, $\epsilon\in\mathbb{R}^+$, there exists a polynomial $P^{MI}_{\kappa,\epsilon}$ of degree $O(\kappa\log(\kappa/\epsilon))$ such that $|P^{MI}_{\kappa,\epsilon}|\leq 1$ in range $[1/\kappa,1]$, which is an $\epsilon/2\kappa$-approximation of the function $f(x)=1/(2\kappa x)$.
\end{lemma}
\textit{Proof:} See Appendix C. in Ref. \cite{Martyn2021}.
\begin{theorem}[Matrix inversion via QSVT]\label{thm: inv}
Let $A$ be a matrix with singular values in the range $[\norm{A}/\kappa_A,\norm{A}]$ for some $\kappa_A\geq 1$, in which $U_A$ is $(\alpha,a,\epsilon)$-block-encoding of $A$ implemented in time $T_A$. Let $\delta\in (0,1]$ such that $\epsilon\leq\delta\norm{A}/(2\kappa_A^2)$, then we can implement a $(2\kappa_A/\norm{A},a+1,\delta)$-block-encoding of $A^+$ in time
\begin{equation*}
    T_{A^+}=O\qty(\frac{\kappa_A\alpha}{\norm{A}}\log\qty(\frac{\kappa_A}{\delta\norm{A}})T_A).
\end{equation*}
\end{theorem}
\textit{Proof:} See Appendix C. in Ref. \cite{Martyn2021}.

Ref. \cite{Chakraborty2023} proposes a quantum algorithm for the Tikhonov regularization in the block-encoding formalism. What we need in particular is their Corollary 33 but with a modified augmented matrix
\begin{equation}\label{eq: augment}
    X_I\equiv\mqty(X&\sqrt{\lambda}I\\0&0),
\end{equation}
which, upon performing $X_I^+=X_I^\dagger\qty(X_IX_I^\dagger)^{-1}$, yields the matrix
\begin{equation}\label{eq: augment_inv}
    X_I^+ = \mqty(X^\dagger\qty(XX^\dagger+\lambda I)^{-1}&0\\0&0),
\end{equation}
containing the Tikhonov-regularized pseudoinverse as its top-left block. 
The time complexity to construct the modified augmented matrix \eqref{eq: augment} is given in the following lemma. 

\begin{lemma}[Tikhonov regularized matrix pseudoinversion]\label{lemma: pinv_lambda}
    Let $\lambda\in\mathbb{R}^+$, $M,N\in\mathbb{N}$. Suppose that we have an $(\alpha_X,a_X,\epsilon_X)$-block encoding of $X\in\mathbb{C}^{M\times N}$ implementable in time $T_X$. Then we can implement an $(\alpha_X+\sqrt{\lambda},a_X,\epsilon_X)$-block encoding of $X_I$ in time $O(T_X)$. Given $\kappa_X$ the condition number of $X$, the condition number of $X_I$ is
    \begin{equation}
        \kappa = \kappa_X\sqrt{\frac{\norm{X}^2+\lambda}{\norm{X}^2+\lambda\kappa_X^2}}.
    \end{equation}
    Then for $\delta\in(0,1]$ such that
    \begin{equation*}
        \epsilon_X\leq\frac{\delta\norm{XX^\dagger+\lambda I}}{32\qty(\alpha+\sqrt{\lambda})\kappa^3\log^3(\kappa/\delta)},
    \end{equation*}
    we will have a $\qty(2\kappa/\qty(\norm{X}+\sqrt{\lambda}),a_X+1,\delta)$-block encoding of $X_I^+$ with time complexity
    \begin{equation*}
        O\qty(\frac{\kappa\alpha_X}{\norm{A}+\sqrt{\lambda}}\log\qty(\frac{\kappa}{\delta})T_X).
    \end{equation*}
\end{lemma}
\textit{Proof:} The augmented matrix in Eq.~\eqref{eq: augment} can be block encoded by virtue of Lemma 18 in Ref. \cite{Chakraborty2023}, but using $\textsc{SWAP}\cdot(I\otimes X)$ as a $(1,1,0)$-block-encoding of $M_B = \mqty(0&1\\0&0)$ to augment $\sqrt{\lambda}I$ in the top-right of the matrix. This augmentation does not change the block-encoding parameters nor the time complexity. Therefore, the block encoding of Eq.~\eqref{eq: augment_inv} satisfies Corollary 33 in Ref. \cite{Chakraborty2023}.

\subsection{Proof of Theorem \ref{thm: W}}



Here, we detail the proof of Theorem \ref{thm: W}.
Suppose that we have a block encoding of $X$, then we can utilize Lemma \ref{lemma: pinv_lambda} to perform the Tikhonov-regularized pseudoinverse of $X$. Now, we have a $\qty(\frac{2\kappa}{\norm{X}+\sqrt{\lambda}},\max(2d+3,t)+1,\delta_X)$-block encoding of $X$, implementable in time $O\qty(\qty(\frac{\kappa}{\norm{X}+\sqrt{\lambda}})\log\qty(\frac{\kappa}{\delta_X})T_\O\sqrt{T})$, where $\delta_X$ is the error associated with the matrix inversion process. Then we perform the multiplication with $Y$ under Lemma \ref{lemma: mult} and obtain a $\qty(\frac{2\sqrt{2}\kappa\|Y\|}{\norm{X}+\sqrt{\lambda}},2\max(2d+3,t)+2,\sqrt{2}\|Y\|\delta_X+\frac{2\kappa\delta_Y}{\norm{X}+\sqrt{\lambda}})$-block encoding of the weight matrix $W$ in time
\begin{equation*}
T_W=O\qty(\qty(\frac{\kappa}{\norm{X}+\sqrt{\lambda}}\log\qty(\frac{\kappa}{\delta_X})+\frac{1}{\|Y\|}\log\qty(\frac{\|Y\|}{\delta_Y}))T_\O\sqrt{T}).
\end{equation*}

Now we can simplify the error of $W$,
\begin{align*}
\delta_W&:= \sqrt{2}\|Y\|\delta_X+\frac{2\kappa\delta_Y}{\norm{X}+\sqrt{\lambda}}\\
&\leq \sqrt{2}\|Y\|\delta_X+\frac{2\kappa\delta_Y}{\norm{X}},
\end{align*}
with the constraint $\delta_X\leq\frac{\delta_W}{2\sqrt{2}\norm{Y}}$ and $\delta_Y\leq\frac{\delta_W}{4\kappa}$ so that $\delta_W\in(0,1]$. Thus, we can rewrite the time complexity $T_W$ to
\begin{equation*}
T_W=O\qty(\qty(\frac{\kappa}{\norm{X}+\sqrt{\lambda}}+\frac{1}{\norm{Y}})\log\qty(\frac{\kappa\norm{Y}}{\delta_W})T_\O\sqrt{T}),
\end{equation*}
as shown in Theorem \ref{thm: W}.

\subsection{Matrix norm inequalities}\label{subsec: matrix_norm}

The operator norm of any matrix $A\in\mathbb{C}^{M\times N}$ can be bounded in terms of the operator norm of matrices $X$ and $Y$ as follows:
\begin{equation}
    \frac{1}{\sqrt{N}}\norm{A}_\infty\leq\norm{A}\leq\sqrt{M}\norm{A}_\infty.
\end{equation}
Since the column elements of matrices $X$ and $Y$ are unit vectors (as they are state vectors), we have $\norm{A}_\infty=\max_{1\leq i\leq M}\sum_{j=1}^N|A_{ij}|=\max_{1\leq i\leq M}\norm{\vb*{A}_i}_1$, where $\vb*{A}_i$ represents the $i$th row of matrix $A$. Additionally, $\norm{\vb*{A}_i}_2\leq\norm{\vb*{A}_i}_1\leq\sqrt{N}\norm{\vb*{A}_i}_2$ holds true if each row of $A$ is a unit vector, resulting in the inequality $1\leq\norm{\vb*{A}_i}_1\leq\sqrt{N}$. Consequently, the inequality for the operator norm of $A$ becomes $\frac{1}{\sqrt{N}}\leq\norm{A}\leq\sqrt{MN}$. In our specific case, with $A=X^T$, $M=4D^2+2D$, and $N=T$, we can deduce that the upper bound of $\norm{X}$ is $O(D\sqrt{T})$, and the lower bound is $\Omega(1/\sqrt{T})$. For matrix $Y$, the upper bound and lower bound become $O(\sqrt{DT})$ and $\Omega(1/\sqrt{T})$, respectively. Referring to the time complexity of $W$, we can rewrite $T_W$, assuming $\lambda=O(1)$, as
\begin{equation}
    T_W = O\qty(\kappa T\log\qty(\frac{\kappa DT}{\delta_W})T_\O).
\end{equation}

In the prediction phase, applying the block-encoded matrix into quantum states requires Lemma \ref{lemma: operation_state}. We can show that the time complexity for applying $W$ is
\begin{equation*}
O\qty(\frac{\kappa\kappa_W}{\norm{X}+\sqrt{\lambda}}\frac{\norm{Y}}{\norm{W}}\log\qty(\frac{\kappa_W}{\delta})T_W+\kappa_WT_{\widetilde \O}). 
\end{equation*}
It commonly knows, from the inequality of matrix norm, that $\norm{Y}=\norm{WX}\leq\norm{W}\norm{X}$. Thus, the complexity is simply written as $O\qty({\kappa\kappa_W}\log\qty(\kappa_W/\delta)T_W+\kappa_WT_{\widetilde \O})$.

\section{Quantum circuit for $m$-time-delay degree-$p$ monomial encoding}\label{sec: time_delay}
\begin{figure*}
    \centering
    \subfigure[]{\includegraphics[width=0.45\textwidth]{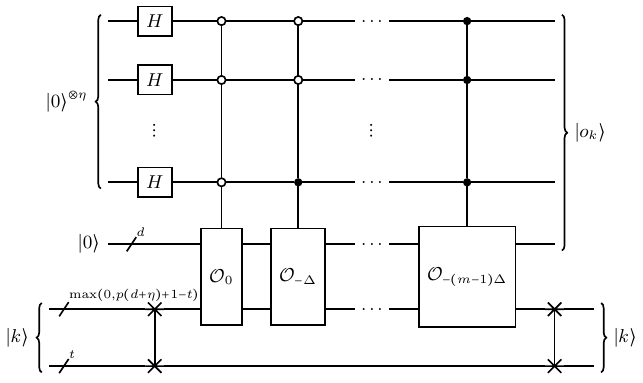}} 
    \subfigure[]{\includegraphics[width=0.45\textwidth]{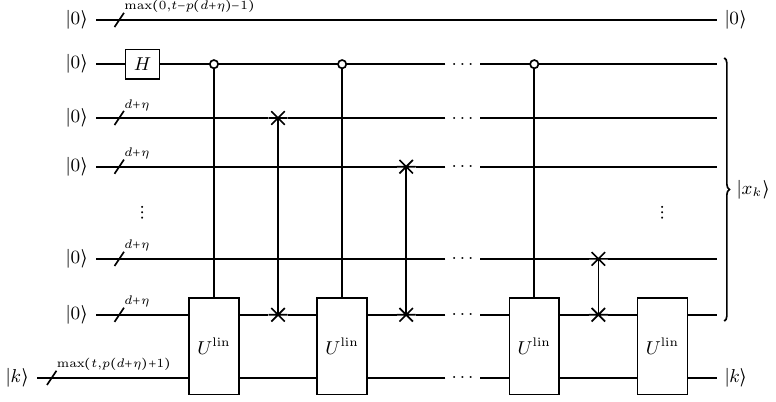}}
    \label{fig: degree_p}
    \caption{Quantum circuit for constructing $m$-time-delay, degree-$p$ feature vectors in general NG-RC. {(a) The quantum circuit $U^\text{lin}$ for the linear part of feature vectors with $m$-delay~\eqref{eq: k-time-delay}, it requires $m$ oracles and $\eta = \log{m}$ additional qubits to construct $\sum_{j=0}^{m-1} \dyad{j}\otimes\mathcal{O}_{-j\Delta}$. (b) $p$ copies of $U^\text{lin}$ are required to construct the degree-$p$ feature vectors.}}
\end{figure*}

As mentioned in the outlook of this work, the NG-RC paradigm encompasses scenarios in which the feature vector is composed of degree-$p$ monomials~\eqref{eq: degree-p} with an $m$-time-delayed $\ket{o_k}$ ~\eqref{eq: k-time-delay}. Here we illustrate a quantum circuit of this encoding in Fig.~\ref{fig: degree_p}(a). When dealing with $m$-time delay data characterized by a step size $\Delta$, we modify the existing quantum circuit $U^\text{lin}$ from Fig.~\ref{fig: roadmap} by incorporating a series of oracles referred to as $\mathcal{O}_{-j\Delta}$, where
\begin{equation}
    \mathcal{O}_{-j\Delta}: \ket{0}^{\otimes d}\ket{k}\mapsto\ket{s_{k-j\Delta}}\ket{k},
\end{equation}
with denoting the number of calls to the oracles by $T_\O$. In this arrangement of oracles, we add other $\eta-1$ Hadamard gates, with $\eta = \log m$, and replace the quantum circuit  $\dyad{0}\otimes\mathcal{O}_0+\dyad{1}\otimes\mathcal{O}_{-1}$ in Fig. \ref{fig: roadmap} by $\sum_{j=0}^{m-1} \dyad{j}\otimes\mathcal{O}_{-j\Delta}$. Since constructing $\ket{o_k}$ involves the use of $m$ oracles, the complexity of this circuit is $O(mT_\O)$.

Now constructing a degree-$p$ feature vector~\eqref{eq: degree-p} requires taking the $p$-fold tensor product of $\ket{o_k}$. Since the no-cloning theorem forbids the creation of $\ket{o_k}^{\otimes p}$ from a single copy of $\ket{o_k}$, this necessitates multiple instances of $U^\text{lin}$ in the quantum circuit.
In particular, we can take the circuit
\begin{equation}
    U^f = \qty(\dyad{0}\otimes \qty(U^\text{lin})^{\otimes p}+\dyad{1}\otimes I^{\otimes p-1}\otimes U^\text{lin})(H\otimes I^{\otimes p}),
\end{equation}
shown in Fig. \ref{fig: degree_p}(b), to be the feature encoder. The overall complexity associated with the creation of $m$-delay and degree-$p$ feature vectors thus equals $O(mpT_\O)$.

\section{Iterative prediction}\label{sec: iterative}

\begin{figure*}[ht]
    \resizebox{\linewidth}{!}{
    \centering
    \includegraphics[scale=1]{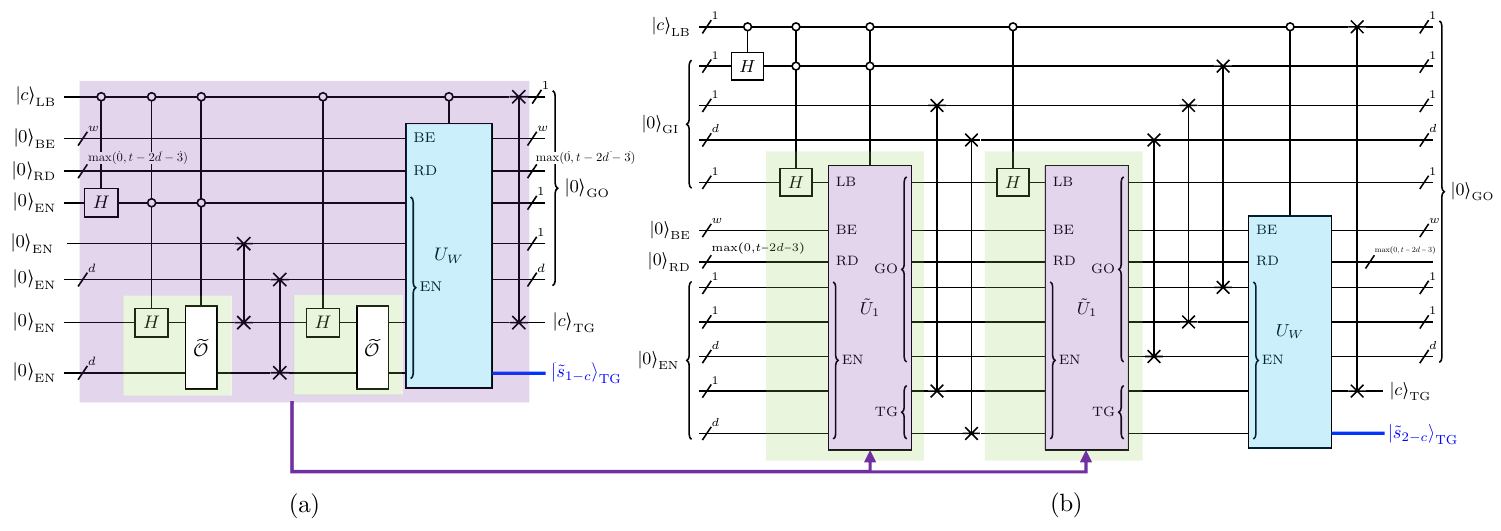}}
    \caption{The recursive quantum circuit for predicting $k=2$ steps into the future.
    In general, at the $j$th level in the recursion ($1\le j\le k$),
    the state $\ket{c}_{\textrm{LB}}$ of the first register LB (stands for ``label") dictates whether the circuit for the $j$th level returns $\ket{\Tilde s_{j-1}}$ (when $c=1$) or $\ket{\Tilde s_{j}}$ (when $c=0$). 
    Thus, to create a linear superposition $\ket{\Tilde{o}_j}=(\ket{0}\ket{\tilde{s}_{j-1}}+\ket{1}\ket{\tilde{s}_{j}})/\sqrt{2}$ for the feature vector, there is always a Hadamard gate applied to the wire going into the LB register (green boxes in quantum circuits).
    Contained in the register BE (``block encoding") are index qubits that label the upper-left block of $U_W$ in the block encoding, whereas qubits in the register RD (``residual") are required for the zero padding to make $W$ square. The EN (``encoder") register takes the data to be encoded into the feature vector. 
    The desired results are contained in the TG (``target") register, whereas the rest are ``garbage output" (GO). The GO registers do not contain any of the relevant result but they remain in the state $\ket{0}$. Thus, we can choose to reuse some of them, and these reused registers are labeled ``garbage input" (GI).
    }
    \label{fig: prediction}
\end{figure*}



In this section, we provide a quantum circuit for the traditional NG-RC approach of predicting $\widetilde{T}$ steps in the future by iteratively predicting one step at a time. 
In this approach the inputs to the prediction phase are past quantum states incorporated via the oracle $\widetilde{\mathcal{O}}: \ket{\tilde{k}}\ket{0}^{\otimes d}\mapsto \ket{\tilde{k}}\ket{\tilde{s}_{-\tilde{k}}}$, where $\tilde{k}=0,1$.
Since the prediction of the state at each time step requires the prediction of the state from the previous two time steps, the predicting circuit then has a recursive structure. Fig. \ref{fig: prediction} shows the circuit that predicts the state $k=2$ steps ahead (Fig. \ref{fig: prediction}(b)), which contains the circuit that predicts one time step ahead (Fig. \ref{fig: prediction}(a)). Note the similarity between the overall structure of the current circuit and that of the circuit for $U^f$ in the training phase, see Fig. \ref{fig: roadmap}. 
For example, the group of gates in the green boxes is responsible for the creation of a linear feature vector, and thus corresponds to the $U^{\textrm{lin}}$'s in the training phase of the algorithm.
The relevant output at the end of the $j$th level 
of the recursion is the prediction $\ket{\tilde{s}_k}= W\ket{\tilde{x}_{k-1}}/\norm{W\ket{\tilde{x}_{k-1}}}_2$ of the state after time step $j$ of the dynamics, obtained from applying $U_W$ to the feature vector $\ket{\tilde x_k}$.
The query complexity of an applying $W$ is $O\qty(\frac{\kappa\kappa_W}{\norm{X}+\sqrt{\lambda}}\frac{\norm{Y}}{\norm{W}}\log\qty(\frac{\kappa_W}{\delta})T_W+\kappa_WT_{\tilde O})$, where $\kappa_W$ is the condition number of $W$, due to the pre-amplification technique for block encoding in Ref. \cite{Chakraborty2023}.

\subsection{Number of ancilla qubits}
The total circuit at the $j$th level represents the unitary operator 
\begin{equation}
    \tilde{U}_j: \ket{c}\ket{0}^{\otimes (w'+2d+3)}\mapsto\ket{0}^{\otimes (w'+d+3)}\ket{c}\ket{\tilde{s}_{j-c}}.
\end{equation}
The upper-left block of $U_W$ that encodes the weight matrix $W$ is a square matrix of dimension $\max(T,8D^2)\times \max(T,8D^2)$, including the zero padding. The block encoding requires in total $w'=w+\max(0,t-2d-3)$ ancilla qubits. The $w$, given in Theorem \ref{thm: W}, is the number of index qubits in the register BE, and $\max(0,t-2d-3)$ is the number of qubits used in the zero padding in the register RD, see Fig. \ref{fig: prediction} and the corresponding caption. 
In total, $w'+d+1+k(d+3)$ qubits are used to predict $k$ time steps.

\subsection{Error propagation}

By allowing imperfections in block encodings, each application of the block-encoded weight matrix $W$ to the feature vector incurs an error, cf. Lemma \ref{lemma: operation_state}. Here we show that, in the iterative prediction scheme, the block-encoding error accumulates exponentially in the number of predicted time steps. However, it is important to note that these errors solely come from the quantum-circuit-level implementation of the QNG-RC, and it is of a fundamentally different kind than the errors in the simulation results displayed in Fig. \ref{fig: QNGRC_sim}. Below, we use the ``prime" symbol to denote the actual state produced in the computation, which can deviate from the ideal state.

Suppose that at the $j=1$ level in the recursion, the state $W\ket{\tilde{x}_0}/\norm{W\ket{\tilde{x}_0}}= \ket{\tilde{s}'_1}$ is $\delta_1$-close to $\ket{\Tilde{s}_1}$, i.e., $\ket{\tilde{s}'_1}$ deviates slightly from the true state $\ket{\Tilde{s}_1}$ with
\begin{equation*}
\norm{\ket{\Tilde{s}_1}-\ket{\Tilde{s}'_1}}_2 \leq \delta_1.
\end{equation*}
(We assume that $\ket{\tilde x_0}$ is error-free.) We can compute an error of the feature vector $\ket{\tilde{x}_1}=\frac{1}{\sqrt{2}}\ket{0}\otimes\ket{\tilde{o}_1}\otimes\ket{\tilde{o}_1}+\frac{1}{\sqrt{2}}\ket{1}\otimes\ket{0}^{\otimes (d+1)}\otimes\ket{\tilde{o}_1}$ in the second step, which is propagated from an error of $\ket{\Tilde{o}'_1}=\frac{1}{\sqrt{2}}\ket{0}\otimes\ket{\tilde{s}'_1}+\frac{1}{\sqrt{2}}\ket{1}\otimes\ket{\tilde{s}_0}$. We know that $\norm{\ket{\tilde{s}_1}-\ket{\tilde{s}'_1}}_2\leq \delta_1$. 
Thus,
\begin{equation}
    \norm{\ket{\tilde{o}_1}-\ket{\tilde{o}'_1}}_2=\norm{\frac{1}{\sqrt{2}}\qty(\ket{0}\otimes\ket{\tilde{s}_1}+\ket{1}\otimes\ket{\tilde{s}_0})-\frac{1}{\sqrt{2}}\qty(\ket{0}\otimes\ket{\tilde{s}'_1}+\ket{1}\otimes\ket{\tilde{s}_0})}_2\leq \frac{\delta_1}{\sqrt{2}}.
\end{equation}
By the triangle inequality, the error in the quadratic tensor product is at least twice the error in the linear part.

\begin{align*}
    \norm{\ket{\tilde{o}_1}\otimes\ket{\tilde{o}_1}-\ket{\tilde{o}'_1}\otimes\ket{\tilde{o}'_1}}_2 
    &= \norm{\ket{\tilde{o}_1}\otimes\ket{\tilde{o}_1}
    - \ket{\tilde{o}'_1}\otimes\ket{\tilde{o}_1}
    + \ket{\tilde{o}'_1}\otimes\ket{\tilde{o}_1}
    -
\ket{\tilde{o}'_1}\otimes\ket{\tilde{o}'_1}}_2 \\
    &= \norm{ (\ket{\tilde{o}_1} - \ket{\tilde{o}'_1}) \otimes \ket{\tilde{o}_1}
    + \ket{\tilde{o}'_1} \otimes (\ket{\tilde{o}_1} - \ket{\tilde{o}'_1})}_2 \\
    &\le 2 \norm{\ket{\tilde{o}_1} - \ket{\tilde{o}'_1}}_2 \le \sqrt{2}\delta_1
\end{align*}
Therefore, the error of the feature vector is bounded by
\begin{align}
    \norm{\ket{\tilde{x}_1}-\ket{\tilde{x}'_1}}_2&=\norm{\frac{1}{\sqrt{2}}\ket{0}\otimes\qty(\ket{\tilde{o}_1}\otimes\ket{\tilde{o}_1}-\ket{\tilde{o}'_1}\otimes\ket{\tilde{o}'_1})+\frac{1}{\sqrt{2}}\ket{1}\otimes\ket{0}^{\otimes (d+1)}\otimes\qty(\ket{\tilde{o}_1}-\ket{\tilde{o}'_1})}_2\\
    &\le 
\frac{1}{\sqrt{2}} \norm{\ket{\tilde{o}_1}\otimes \ket{\tilde{o}_1} - \ket{\tilde{o}'_1}\otimes \ket{\tilde{o}'_1}}_2 + \frac{1}{\sqrt{2}} \norm{\ket{\tilde{o}_1} - \ket{\tilde{o}'_1}}_2
    \le \frac{3\delta_1}{2}.
\end{align}

Ref. \cite{Chakraborty2023} also shows that the error of the application of a block-encoded matrix on an erroneous quantum state can be computed by Lemma \ref{lemma: error_propagation}. 

\begin{lemma}[Error propagation]\label{lemma: error_propagation}
    Suppose that $\ket{b'}$ is $\delta/2\kappa$-close to $\ket{b}$, where $\kappa$ is the condition number of a block-encoded matrix $A$ such that its singular values lie in $[\norm{A}/\kappa,\norm{A}]$. A quantum state $\ket{\psi}$ that is $\delta/2$-close to $A\ket{b'}/\norm{A\ket{b'}}_2$ will be $\delta$-close to $A\ket{b}/\norm{A\ket{b}}_2$.
\end{lemma}

According to Lemma \ref{lemma: error_propagation}, we will have, in our work, an inequality relation
\begin{equation}
    \frac{3}{2}\delta_1 \leq\frac{\delta_2}{2\kappa_W}
\end{equation}
where $\delta_2\in(0,1]$ is an error of $\ket{\Tilde{s}_2}$ such that is $\delta_2$-close to $W\ket{\tilde{x}_1}/\norm{W\ket{\tilde{x}_1}}_2$.

The error from block encoding of the $k$th prediction vector $\ket{\tilde{s}_k}$ can be computed from the error propagation of $\ket{\tilde{x}_{k-1}}$ and successive applications of the blocked-encoded unitary $U_W$:
\begin{equation}
    \norm{\ket{\tilde{o}_{k-1}}-\ket{\tilde{o}'_{k-1}}}_2=\norm{\frac{1}{\sqrt{2}}\qty(\ket{0}\otimes\ket{\tilde{s}_{k-1}}+\ket{1}\otimes\ket{\tilde{s}_{k-2}})-\frac{1}{\sqrt{2}}\qty(\ket{0}\otimes\ket{\tilde{s}'_{k-1}}+\ket{1}\otimes\ket{\tilde{s}'_{k-2}})}_2\leq \frac{1}{\sqrt{2}}(\delta_{k-1}+\delta_{k-2})
\end{equation}
and
\begin{align*}
    \norm{\ket{\tilde{o}_{k-1}}\otimes\ket{\tilde{o}_{k-1}}-\ket{\tilde{o}'_{k-1}}\otimes\ket{\tilde{o}'_{k-1}}}_2
    &\leq\sqrt{2}(\delta_{k-1}+\delta_{k-2})
\end{align*}
Therefore, the error of $\ket{\Tilde{x}_{k-1}}$ is
\begin{equation}
    \norm{\ket{\tilde{x}_{k-1}}-\ket{\tilde{x}'_{k-1}}}_2\leq \frac{3}{2}(\delta_{k-1}+\delta_{k-2}).
\end{equation}
and a recurrence relation for the error propagation  can be written as
\begin{equation}\label{eq: error_propagate_k}
    \frac{3}{2}(\delta_{k-1}+\delta_{k-2}) \leq\frac{\delta_k}{2\kappa_W},
\end{equation}
where $\delta_k\in(0,1]$, 
implying that $\delta_j=O\qty(3^{-j}\kappa_W^{j-k}\delta_k)$. That is, the error accumulates exponentially.

\section{NG-RC for a quantum chaotic system}\label{sec: Sup-TiltedField}
\begin{figure}
    \centering
    \includegraphics[scale=0.8]{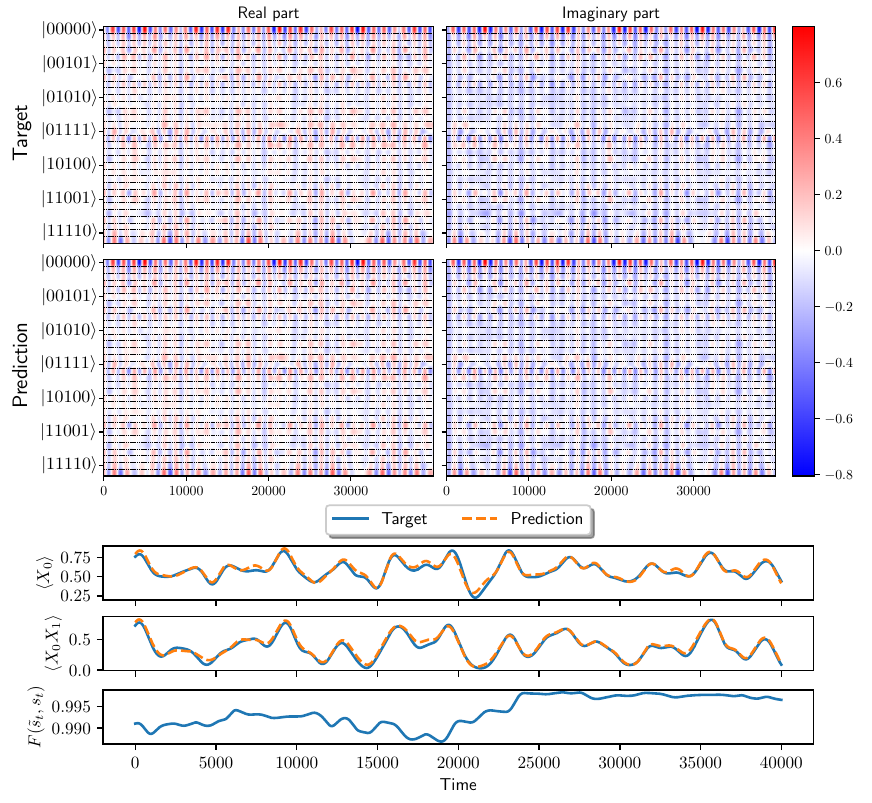}
    \caption{The performances of the NG-RC in predicting the future \emph{unseen} states of a five-qubit tilted-field Ising model~\eqref{eq: tilted} in the \emph{chaotic} phase with $J = 1$, $h = 1$, and $\theta = 15\pi/32$ across $\widetilde T = 4 \times 10^4$ future time steps employing the skipping-ahead method with a time skip of $\tau=10^6$ steps.
    $T=2\times 10^4$ steps of the time evolution with the same step size of $\Delta t=1/(200E_{\textrm{max}})$ are used to train the NG-RC, with the regularization parameter $\lambda = 0$. Our benchmark targets are the real and the imaginary parts of all amplitudes in the computational basis (top). The comparisons of the observables $\expval{X_0}$, $\expval{X_0X_1}$, and the fidelity $F(\tilde{s}_t,s_t):=|\braket{\tilde{s}_t}{s_t}|$ between the target and the predicted states are also shown (bottom).} 
    \label{fig: tilted_Ising}
\end{figure}

The original NG-RC has been applied to successfully predict classical chaotic time series; we now explore the capability of NG-RC in predicting a full many-body quantum chaotic dynamics. One example of quantum chaos arises in the tilted-field Ising model with an open boundary condition, which is known for its non-integrability and chaos for specific tilting angles of an external magnetic field \cite{TFIM_2007}. The Hamiltonian of the system can be written as
\begin{equation}\label{eq: tilted}
    \mathcal{H}_\text{tilted} =J\sum_{i=1}^{d-1} Z_{i}Z_{i+1} + h\sum_{i=1}^d \qty(X_{i}\sin\theta  + Z_{i}\cos \theta),
\end{equation}
such that the signature of quantum chaos appears when the nearest-neighbour level-spacing distribution of the Hamiltonian eigen-energies resembles Wigner-Dyson statistics \cite{AP_2016_Chaos}. We focus on $J = 1$, $h = 1$, and $\theta = 15\pi/32$, so that the level-spacing distribution approaches Wigner-Dyson distribution even when $d=5$. In the prediction phase, the inputs into the trained NG-RC (with $\tau = 10^6$) are obtained from 
the initially uniform superposition state $\frac{1}{2^{d/2}}\sum_{i=0}^{2^d-1}\ket{i}$ that undergo a long burn-in period. Even when the observables appear chaotic, Fig.~\ref{fig: tilted_Ising} shows an excellent prediction accuracy of the system's evolution with $\tau$ time steps ahead into the future.

\end{document}